\documentclass[prd,nofootinbib,floats,superscriptaddress,eqsecnum,tightenlines, showpacs,11pt]{revtex4}

\usepackage{enumerate}
\usepackage{bbold, bm}
\usepackage{graphicx}
\usepackage{subfigure}
\usepackage{showlabels}

\usepackage{array}
\usepackage{setspace}

\usepackage{hyperref}
\usepackage{amsmath,amssymb,amsfonts,amsthm,latexsym,stmaryrd}

\def\be{\begin{eqnarray}}
\def\ee{\end{eqnarray}}
\def\sgn{\mathrm{sgn}}

\def\l{\ell}
\def\w{\bm{\omega}}
\def\z{\bm{z}}
\def\A{\bm{A}}
\def\E{\bm{E}}
\def\K{\bm{K}}

\def\D{\bm{D}}
\def\J{\bm{J}}
\def\L{\bm{L}}
\def\S{\bm{S}}
\def\r{\bm{r}}

\def\t{\bm{\tau}}
\def\X{\bm{X}}

\def\s{\bm{\sigma}}

\def\rd{\mathrm{d}}

\newtheorem{definition}{Definition}

\newcommand{\SU}{\text{SU}}

\newcommand{\su}{\mathfrak{su}}

\newcommand{\Tr}{\mbox{Tr}}

\begin{document}

\title{Spinning geometry = Twisted geometry}

\author{Laurent Freidel}
\email{lfreidel@perimeterinstitute.ca}
\affiliation{Perimeter Institute for Theoretical Physics\\
31 Caroline St. N, N2L 2Y5, Waterloo ON, Canada}\date{\today}

\author{Jonathan Ziprick}
\email{jziprick@perimeterinstitute.ca}
\affiliation{Perimeter Institute for Theoretical Physics\\
31 Caroline St. N, N2L 2Y5, Waterloo ON, Canada}
\affiliation{Department of Physics, University of Waterloo\\
Waterloo, Ontario N2L 3G1, Canada}

\begin{abstract}
It is well known that the $\SU(2)$-gauge invariant phase space of loop gravity can be represented in terms of twisted geometries. These are piecewise-linear-flat geometries obtained by gluing together polyhedra, but the resulting geometries are not continuous across the faces.
Here we show that this phase space can also be represented by continuous, piecewise-flat three-geometries called spinning geometries.
These are composed of metric-flat three-cells glued together consistently.
The geometry of each cell and the manner in which they are glued is compatible with the choice of fluxes and holonomies.
 We first remark that the fluxes provide each edge with an angular momentum.
By studying the piecewise-flat geometries which minimize edge lengths, we show that these angular momenta can be literally interpreted as the spin of the edges: the geometries of all edges are necessarily helices.
We also show that the compatibility of the gluing maps with the holonomy data results in the same conclusion.
This shows that a spinning geometry  represents a way to glue together the three-cells of a twisted geometry to form a continuous geometry which represents a point in the loop gravity phase space.
\end{abstract}

\maketitle

\section{Introduction}
In the quantum field theory of massive particles one defines the theory within a truncation scheme,
restricting first the definition of asymptotic states to a finite number $n$ of particles, then defining the amplitudes recursively between an arbitrary number of in and out particle states.
What this truncation achieves is the possibility to organize the theory in terms of Fock states so that we can deal with finite-dimensional Hilbert spaces at each energy level. It also emphasizes a basis of states that possesses a strong classical interpretation: the particle. This truncation is very different from an approximation scheme such as a discretization. It is not supposed to be
an approximate description of a continuum theory that needs some continuum limit. It is supposed to be an {\it exact} description of the continuum theory restricted to a particularly convenient and finite basis of states.
The full theory lies in the knowledge of all amplitudes, not in some continuum limit.
There are some limiting procedures to be taken which are associated with the reorganization of coupling constants, following from the existence of naive divergences. These divergences are in some sense welcome since they give us strong clues about spacetime locality.

Interestingly, loop quantum gravity and spin foams possess the same set of ingredients.
In loop quantum gravity \cite{Thiemann-book,Rovelli-book}, a truncation is needed in order to define the kinematical Hilbert space.
One truncates the theory by looking first at spin networks states that are supported on a {\it finite} graph $\Gamma$ embedded in space. Then one shows that this leads to finite dimensional Hilbert spaces and discrete spectra.
Spin foams aim at computing all possible amplitudes between such states supported on finite graphs.
Also, there exists interesting naive divergences in these amplitudes. These divergences should be welcome and renormalized; they have been recognized to be related to the existence of spacetime diffeomorphism symmetry \cite{FL}.

Let us emphasize that this point of view is key to understanding the program and is often the source of
much confusion in the recent literature.
If one thinks of loop gravity as a truncation rather than a discretization (or approximation), one should not try to take a naive continuum limit of it. One should instead find a proper way to understand and deal with the reorganization of infinities, and understand the intertwining of these infinities with spacetime diffeomorphism.

It is important to understand the classical nature of the truncation in order to decide whether loop gravity should be treated as a truncation or as a discretization.
In the field theory case, the truncation is associated with particles and can be expressed in terms of classical field configurations with compact topology.
Therefore, the central question we  want to investigate here, is whether it is possible to assign to
the loop gravity truncation a classical meaning in terms of acceptable three-geometries.
Loop gravity is after all a theory of quantum gravity, based on the quantization of a phase space which is interpreted as the cotangent bundle over the space of three-geometries. So the question is whether we can assign to each truncated Hilbert space ${\cal H}_{\Gamma}$ a truncated phase spaces $P_{\Gamma}$, and whether we can understand these truncated phase space $P_{\Gamma}$ to be associated with a classical, albeit discrete, geometry.

There has recently been a great deal of progress on these two questions.
Concerning the first point let us emphasize that contrarily to a naive belief, it is always possible to express quantum states and amplitudes in terms of classical entities. This is the spirit of Feynman path integral quantization which achieves quantization purely in terms of classical entities. On the other hand this is also the spirit of geometrical quantization which emphasizes that one can always chose a basis of states which is coherent, so that states may be labeled by classical phase space points. These two points of view are complementary to each other. A famous example is the spin of a spinning particle which can be understood as an additional classical and compact degree of freedom (see \cite{spin} for different derivations).

This analysis led to the introduction of the discrete holonomy-flux phase space $P_{\Gamma}$: its variables are given by
elements $(\bm{X}_{cc'},h_{cc'})\in \mathfrak{su}(2)\times \mathrm{SU}(2)$ for each pair of nodes $c$ joined by a link in $\Gamma$, subject to certain relations\footnote{Given by
$
h_{c'c}=h_{cc'}^{-1},\qquad \bm{X}_{c'c} = - h_{c'c}\bm{X}_{cc'}h_{cc'}.
$}.
Since $T^{*}\mathrm{SU}(2)=  \mathfrak{su}(2)\times \mathrm{SU}(2)$, this space is naturally a phase space. The left over property of the quantum geometry resides in the fact that the flux variables $X_{cc'}$ do not Poisson commute.

The second question led to the introduction of twisted geometries \cite{DR}.
In \cite{LSpeziale1} it was shown that the truncated loop gravity phase space $P_{\Gamma}$ possesses a natural geometrical interpretation in terms of discrete geometries.
This phase space was shown to be understood as the gluing of convex polyhedra \cite{FLivine,SBianchi} along their faces, leading to a piecewise-linear-flat\footnote{Here `flat' implies that the metric associated to each polyhedron is flat, while `linear' refers to the fact that gluing maps have to be linear which implies  the flatness of the induced metric on each face.} but discontinuous geometry.
The relationship between twisted geometries and  twistors has also been developed \cite{LSpeziale2,Speziale}.
Recently an important development \cite{Haggard} showed that twisted geometries admit a torsionless connection.
This shows that twisted geometries are a natural generalization of Regge geometries.
As a confirmation, the analog of the Regge action for twisted geometries have been found \cite{FJeff} and shown to appear in the asymptotics of $15$j-symbols. The geometrical side of these results can be summarized by the statement:
{\it A twisted geometry is a discontinuous, piecewise-linear-flat manifold which possesses a torsionless connection.}

Despite the success of this approach, one drawback lies in the fact that the geometries obtained are discontinuous across the faces of the polyhedra.
Indeed, the shape of each face appears differently from the perspective of each polyhedra that shares it, and it is therefore not possible to assign a common length to the edges of this geometry.
This is in sharp contrast with a Regge geometry where edge lengths agree along faces that are glued together. This is an issue if one wants to have a well defined notion of frame fields, and it has motivated some authors to impose on the twisted geometries the Regge constraints \cite{Bianca}. However, by doing so we lose the link with the loop gravity phase space, the discreteness of geometrical observables and the power of spin foam quantization.

In order to reconcile this tension a new approach was developed in \cite{FGZ} (see \cite{Bianchi1} for an earlier work in this direction).
Here it was shown that it is possible to relate directly the truncated loop gravity phase space to the full gravity phase space by imposing some flatness constraints adapted to the spin foam graph and its dual.
It was also argued that this implies twisted geometries can be interpreted as {\it continuous}, piecewise-flat geometries carrying torsion.
Here we show that these expectation are satisfied. We introduce in what follows the notion of a {\it spinning geometry}:
{\it a continuous, piecewise-flat geometry with a piecewise-torsionless connection.}

In other words, a spinning geometry carries discrete curvature {\it and} torsion along the edges of the cellular complex, but it is continuous.
We can now appreciate the differences between spinning geometries and twisted geometries:
one is continuous while the other is {\it dis}continuous;
one is piecewise-flat, the other is piecewise-{\it linear}-flat,
one is torsion{\it full}, the other is torsion{\it less}. A summary of this comparison is presented in table \ref{table}.
In a spinning geometry, the curvature and torsion both vanish inside three-cells.
One of the key differences that explains the existence of spinning geometries is the relaxation of the demand that the geometry be linear. If we have a linear geometry it is then necessarily flat, but the converse is not true. There exist flat geometries that are not obtained by linear gluing. An example of such a space as we will see is an helicoidal space obtained by gluing a wedge supported on an helix with a general Poincar\'e transformation admitting a translational component along the wedge.

\begin{table}
{\renewcommand{\arraystretch}{1.2}
\begin{tabular}{c@{\qquad}|@{\qquad}c@{\qquad}c@{\qquad}c}
Geometry & 3-metric & faces & edges\\
\hline
Regge & continuous & flat & torsionless \\
Twisted & discontinuous & flat & torsionless\\
Spinning & continuous & curved & torsionfull
\end{tabular}}
\caption{Regge, twisted and spinning geometries are cellular spaces composed of flat, torsionless three-cells. This table contrasts their differences.}
\label{table}
\end{table}

The main result presented in this paper is the equivalence between the truncated loop gravity phase space $P_{\Gamma}$ and the spinning geometries. This implies an equivalence between spinning geometries and twisted geometries. They represent the same object, and we can therefore think of a twisted geometry as a regular, continuous spinning geometry.

For completeness let us characterize Regge geometries in a similar way.
{\it A Regge geometry is a continuous, piecewise-linear-flat manifold which possesses a torsionless connection.}
We can see now that a Regge geometry lies at the intersection of a spinning geometry and a twisted geometry.
It is continuous, linear, flat and torsionless. In that sense both spinning and twisted geometries are natural generalizations of Regge geometries.
The obvious advantage of the spinning geometries is that they are continuous and do not require an extension of what we demand geometries to be in gravity.

%%%%%%%%%%%%%%%%%%%%%%%%%%%%%%%%%%%%%%%%%%%%%%%%%%%%%%%%%%%%%%%%%%%%%%%%%%%%%%%%%%%%%%%%%%%%%%%%%%%%%%%%

This extension is possible since we allow torsion to be non-vanishing along the edges within cell boundaries.
This puts on the same footing torsion and curvature, which is also supported along these edges. The fact that we allow torsion in the phase space should not come as a surprised since the introduction of the Immirzi parameter in loop gravity amounts to allowing torsion to fluctuate at the quantum level \cite{LArtem}.
It is therefore welcomed to have a formalism that allows such configurations in its phase space.

In the following we first recall some definitions concerning cell complexes, and we define and study regular, flat, cellular geometries, considering certain deformations of the three-cells. We also provide a decomposition of the fluxes in terms of an angular momentum associated with each edge. This decomposition, which follows from the Gauss constraint, is at the core of our geometrical picture.
We then embark in minimizing the lengths of the one-dimensional links in the three-cell boundaries
under a constraint which fixes the angular momenta associated with edges.
We find that generally the edge has to have the shape of a helix.
We show how to relate explicitly the parameters of the helix to the edge angular momenta.
Finally we show that these three-cells with helical edges in their boundaries are glued together to form a continuous geometry.
The parameters of the helices are related to how the cells are glued together.

This article is essentially an exercise in geometry and requires minimal background knowledge of loop gravity, other than definitions of the geometries in question. Let us begin the analysis starting with these definitions in the next section.

\section{Definitions}
\label{def}
In this section we provide definitions of the twisted, flat-cell and Regge geometries. This discussion is relevant for the calculations which follow, but also serves to explain the relationship between these geometries.

We denote in bold letters an element $\A \in \su(2)$, which can be identified with a vector in $\mathbb{R}^3$ through $A^i = -2 \Tr (\bm{A} \t^i)$, where $\t^i$ is an $\su(2)$ basis given by $-i/2$ times the Pauli matrices. This basis satisfies the algebra $[\t_{i},\t_{j}]=\epsilon_{ijk}\t^{k}$.
We will use also the vector notation of a dot-product for a trace $\bm{A} \cdot \bm{B} \equiv -2 \Tr (\bm{AB})$, and a cross-product for an $\su(2)$ commutator $\bm{A} \times \bm{B} \equiv [\bm{A},\bm{B}]$. A magnitude is denoted by dropping the bold font $A \equiv |\A|$, and a unit vector is denoted by a hat $\hat{\A} \equiv \A / A$.

\subsection{Twisted geometry}
A twisted geometry is a set of polyhedra-shaped three-cells $c$ which are `glued' together along faces $f$ according to certain rules. To each face oriented outwardly with respect to the cell $c$, we attribute an $\su(2)$-valued flux $\X_f$ and an $\SU(2)$-valued holonomy $h_f$ (see \cite{newflux} for earlier or related discussion).
%Consider a single cell $c$. If we assume the geometry of $c$ is linear and flat, then the fluxes associated to each face define a polyhedron.
The face areas are given by the magnitudes $X_f$ and the orientations are given by outward pointing unit vectors $\hat{\X}_f$ which are normal to the faces. In order that each polyhedron is closed, the fluxes satisfy a closure relation:
\be
\label{closure}
\sum_f \X_f = 0,
\ee
where the sum is over all the faces of the polyhedron. By a classical theorem of Minkowski \cite{Minkowski}, the reverse is true: A set of flux operators satisfying the closure condition defines a polyhedron.

A face $f_{cc'}$ on a cell $c$ is glued to a face $f_{c'c}$ on a neighbouring cell $c'$. Each face possesses an orientation such that $f_{cc'}$ is oriented oppositely to $f_{c'c}$. Holonomies come into play for gluing polyhedra together. For faces that are glued together the areas are the same, i.e. $X_{cc'} = X_{c'c}$, however the orientation is generally different. A consistent gluing requires that:
\be
\label{gluing}
\X_{c'c} = -h_{{c c^\prime}}^{-1} \X_{cc'} h_{{c c^\prime}}.
\ee
The non-zero extrinsic curvature at the intersections between cells is manifest in these gluing conditions.

A twisted geometry does not form a continuous geometry. This is because the shape of a face $f_{cc'}$ that comes from the polyhedral geometry of $c$, is in general  different than the shape of $f_{c'c}$ which it is glued to. For example, $f_{cc'}$ may have three sides while $f_{c'c}$ can have four. Even if each face possesses the same number of sides, the faces still have different shapes in general, so that the boundary links of each face do not match. In order to form a continuous geometry, one may impose conditions \cite{Bianca} which force the shapes of faces to match when they are glued together. These extra conditions restrict the degrees of freedom in the loop gravity phase space and reduce a twisted geometry to a Regge geometry.
We now explore another option which provides continuous geometries without having to reduce the number of degrees of freedom.

\subsection{Cell complex and flat-cell geometry}
\label{flatdef}
A {\it Regge geometry} is defined as a piecewise-linear-flat cellular complex \cite{Regge}.
This means that it is obtained by gluing together polyhedra along their faces with piecewise-linear maps \cite{Book1}.
A {\it spinning geometry} is defined as a piecewise-flat, regular cellular complex.
This is a geometry which is obtained by gluing cells homeomorphic to polyhedra, but we relax the condition of the
gluing maps to be piecewise-linear. We demand instead that the resulting geometry is flat.
In order to understand the generalization from Regge geometry to spinning geometry we need to review the definition
of a cellular complex.

Here we shall work with a toplogical space that is a generalization of a simplicial complex, referred to as a CW complex $\Delta$ where the `C' denotes that the space is `closure-finite' and the `W' is for `weak topology' \cite{Book2}. The idea is to obtain an $n$-dimensional space by gluing together a collection of $n$-dimensional cells in a non-linear manner so that the cells are generally not simplices. In constructing such a space one needs to work with cells of every dimension $i = 0, \cdots, n$, and in the following definitions we shall denote by $c_i$ a cell of dimension $i$. Each $c_i$ is an open ball $\mathring{B}_{i}$, having a boundary $\partial c_i$ that is topologically equivalent to the $(i-1)$-sphere $S^{i-1}$. We denote by $\bar{c}_{i}$ the closure of the cell $c_{i}$, so that the boundary is given by $\partial c_{i}= \bar{c}_{i} \backslash c_{i}$.

The boundary of a cell $\partial c_{i+1}$ is composed of cells with dimension $0, \cdots, i$, and the union of these boundaries over the entire complex provides a definition of an important structure called the $i$\textit{-skeleton}. Within a CW complex, the $i$-skeleton ${\Delta}_{i}$ is defined recursively by gluing a disjoint union of $i$-dimensional cells $c_i$ to the skeleton of lower dimension ${\Delta}_{i-1}$. In other words, one begins with a collection of $0$-dimensional cells (points) $c_0$ defining a $0$-skeleton, ane obtains from this the $1$-skeleton $\Delta_1$ by connecting the points $c_0$ with one-dimensional edges $c_1$. Taking then the disjoint union of $\Delta_1$ with two-dimensional faces yields the two-skeleton $\Delta_2$, and this process is repeated to define the $i$-skeleton of any dimension $i<n$ within an $n$-dimensional CW complex. Let us now be more precise about how this is done. Suppose that the $(i-1)$-skeleton $\Delta_{i-1}$ of a CW complex is given. We introduce {\it gluing maps} $s^{i}$ and define $\Delta_{i}$ by gluing $i$-dimensional cells to $\Delta_{i-1}$:
\be
s^{i}: \partial c_{i} \to \Delta_{i-1},\qquad \Delta_{i}\equiv \left(\Delta_{i-1}\bigsqcup_{c_i} c_{i}\right)/ \sim,
\ee
where the quotient by $\sim$ denotes the identification
provided by the gluing maps: given $x\in \partial c_{i}$, $y \in \Delta_{i-1}$ we say that $ x\sim y$ if
$s^{i}(x)=y$. This formula means that we obtain $\Delta_{i}$ by quotienting the disjoint union of 
$\Delta_{i-1}$ and $c_{i}$ by the identification relation provided by the gluing maps $s^{i}$.
In this way we can start with a set of points $\Delta_0$ and build up to dimension $i$ recursively as described above.
Note that $\bar{c}_{i}$ is itself included in $ \Delta_{i}$ and
under this inclusion  it  becomes itself an {\it elementary} cell complex whose boundary can be decomposed into cells of different dimensions.

Since we are interested specifically in dimension $3$ we will denote the three-dimensional open cells by $c$, the two-dimensional open cells (called faces) by $f$, the one dimensional open cells (the edges or links) by $\l$, and the zero dimensional cells (the nodes) by $n$.
In the following $(c, f, \l)$ always denote  open cells, their closures are denoted by $(\bar{c},\bar{f},\bar{l})$ and their boundaries are denoted by, for example, $\partial c = \bar{c}\backslash c$.

A CW complex can be a very general object. Here we are going to study a subclass of CW complexes that we call regular (see \cite{Kirillov} for a related discussion in the  piecewise linear context).
\begin{definition}
A regular three-dimensional cellular space $\Delta$ is a collection of three-dimensional cells  $c$,
glued together. We demand that: \\
1) The closure of each cell is diffeomorphic to a convex polyhedra $P_{c} \subset \mathbb{R}^{3}$;
the diffeomorphism is denoted $\psi_c: \bar{c} \to P_{c}$.
We denote by $(f_{c},\l_{c},n_{c})\subset \partial c$, the inverse images of (respectively) the faces, edges and nodes of the boundary of $P_{c}$.\\
2) There exist invertible gluing maps for the unique pair of cells that are glued along $f$:
\be
\begin{array}{rl}
s_{cc'}: & \bar{f}_{c} \to \bar{f}_{c'}. \\
% & \l_{c} \to \l_{c'} \\
% & n_{c} \to n_{c'}
\end{array}
\ee
Moreover the restriction of these maps to the boundary of the face $\partial f_{c}$ are invertible maps onto $\partial f_{c'}$.
%There also exist invertible gluing maps  $s^{\l}_{cc'}: \l_{c} \to l_{c'}$, for the each pair of of cells that share a common edge.
The three-dimensional cell complex is defined as the quotient space $\bigsqcup_{c}c /\sim$ where $x\in \bar{f}_{c}$ is equivalent
to $y \in \bar{f}_{c'}$ when $s_{cc'}(x)=y$.
\end{definition}

%In the following we will call a {\it regular n-complex} a CW complex such that no more than two the top dimensional cells are  attached to a dimension $n-1$ cell.
From the above definition we can develop the notion of a three-dimensional piecewise-flat cellular space.
\begin{definition}
A three-dimensional piecewise-flat cellular space is a regular complex ${\Delta}$ together with a (not necessarily piecewise-linear) embedding of $c$ into $\mathbb{R}^{3}$. That is we have a set of injective maps:
\be
\bm{z}^{c}: \bar{c} \to \mathbb{R}^{3}.
%\qquad z^{f}: \bar{f} \to \mathbb{R}^{3},\qquad z^{\l}: \bar{\l} \to \mathbb{R}^{3}.
\ee
These maps define a flat metric $(g^{c})_{\mu \nu}:= \partial_{\mu} \z^c \cdot \partial_{\nu} \z^c$ on each cell $c$.
We demand these metrics to be compatible with the gluing maps:
\be\label{comp}
(s_{cc'})^{*} g_{c'}(x) =  g_{c}(x), \qquad \forall x \in \bar{f}_{c}.
\ee
\end{definition}
The key point to appreciate here is that even if $\bar{c}$ is equipped with a flat metric, it does not have to coincide with the polyhedral metric on $\psi_c(\bar{c})= P_{c}$. In particular we {\it do not} assume that
the induced metric on the faces of $\bar{c}$ is flat. That is we allow an arbitrary value of the extrinsic curvature tensor on the faces of $\bar{c}$.

The condition of compatibility (\ref{comp}) of the metric with the gluing maps can be expressed more explicitly in terms of the flat coordinates $\z^{c}$. Indeed it implies that the coordinates $\z^{c}$ and $\z^{c'}$ are related by a Poincar\'e transformation when evaluated on the boundary of the cell.

Consider two neighbouring cells $c$ and $c^\prime$. From now on we will denote by $f_{c c^\prime}$ the face shared by $c$ and $c'$ viewed from $c$, and by $f_{ c^\prime c} =s_{cc'}(f_{c c^\prime})$ the same face as seen from $c'$. We assign to each face their outward orientation so that $f_{cc'}$ and $f_{c'c }$ have opposite orientations. Under the gluing maps we obtain a continuous metric on cell faces with the requirement:
\be
\label{z12i}
\rd \bm{z}^{c^\prime}(s_{cc'}(x) ) = h_{{c c^\prime}}^{-1} \rd \bm{z}^{c}(x) h_{{c c^\prime}}, \hspace{0.5in} \forall x \in \bar{f}_{cc'},
\ee
where $h_{c c^\prime} \in \SU(2)$ is a group element associated to the face $f_{cc'}$ such that $h_{c'c} = h_{cc'}^{-1}$.
This implies that the coordinate functions are related by a Poincar\'e transformation:
\be
\label{z12}
\bm{z}^{c^\prime}(s_{cc'}(x) ) = h_{{c c^\prime}}^{-1} (\bm{z}^{c}(x)+\bm{a}_{c c^\prime}) h_{{c c^\prime}}, \hspace{0.5in} \forall x \in \bar{f}_{cc'},
\ee
where $\bm{a}_{c c^\prime} \in \su(2)$ is an algebra element corresponding to a translation. While (\ref{z12i}) implies (\ref{z12}), notice that the reverse is not true\footnote{If $\z^c$ and $\z^{c'}$ are related by a Poincar\'e transformation, only the differential tangential to the face satisfies (\ref{z12i}). This equation also expresses that the derivative
normal to the face has to be continuous. This condition gives us information about the shape of the faces as embedded in the three-cells. We will postpone the detailed study of these conditions, since here we focus on the shapes of the edges.
}.
This suggests the following definitions:
\begin{definition}
A {\rm{Regge}} geometry is a piecewise-flat cellular space such that the induced metric on all of the faces is flat.
A {\rm{spinning}} geometry is a piecewise-flat cellular space such that the image of edges of $\bar{c}$ by $z^{c}$ are helices.
\end{definition}
We will study in sections \ref{mom} and \ref{min} how the helical shape of the edges in a spinning geometry arise from the definitions we have given.

\subsection{Geometric representations of the holonomy-flux phase space}

In order to keep this article self-contained, we now present a brief description of the holonomy-flux phase space of loop gravity. Since both spinning and twisted geometries are different representations of this same space, we will see that these two geometries are in fact isomorphic to each other. For further details on how the holonomy-flux phase space is represented as a twisted geometry we refer the reader to \cite{LSpeziale1}, and for more details on how this phase space relates to spinning (or flat cell) geometries we refer the reader to \cite{FGZ}, specifically section IV.B.

The holonomy-flux phase space $P_\Gamma$ is defined upon a graph $\Gamma$, which is simply a network of one-dimensional edges $e$ connected at their endpoints, or vertices $v$. Each edge is assigned a holonomy $h_e \in \SU(2)$ and a flux $\X_e \in \su(2)$, so that the phase space associated to a single edge is the cotangent bundle $T^* \SU(2)$. Taking the direct product over all of the edges, one obtains the graph phase space:
\be
P_\Gamma\equiv\underset{e}{\times}T^*\SU(2)_e.
\ee
%The Poisson algebra associated to this phase space is given by:
%\be\label{poisson algebra2}
%\big\lbrace X^i_e,X^j_{e'}\big\rbrace=\delta_{ee'}\epsilon^{ij}_{~~k}X^k_e,\qquad\big\lbrace X^i_e,h_{e'}\big\rbrace=-\delta_{ee'}\t^i h_e,\qquad\big\lbrace h_e,h_{e'}\big\rbrace=0.
%\ee
The fluxes are subject to a discrete Gauss law at each vertex of the graph:
\be
\sum_{e \ni v} \X_e = 0,
\ee
where the sum is over all edges $e$ which intersect at the vertex $v$.
This constraint generates $\SU(2)$ gauge transformations which act at the vertices of the graph.
%\be\label{SU(2)}
%g_v\triangleright h_e=g_{s(e)}h_eg_{t(e)}^{-1},\qquad\qquad g_v\triangleright \X_e=g_{s(e)} \X_eg_{s(e)}^{-1},
%\ee
%where $s(e)$ (resp. $t(e)$) denotes the starting (resp. terminal) vertex of $e$, and the $g_v \in \SU(2)$ are associated to the vertices.

Using symplectic reduction, one can incorporate the Gauss law into the holonomy-flux phase space by: 1) requiring that the fluxes satisfy the constraint at each vertex; 2) identifying sets of data that are related by $\SU(2)$ gauge transformations. We use a double slash notation to denote this two-step process, and define the $\SU(2)$-gauge invariant phase space:
\be
P^G_\Gamma \equiv P_\Gamma \sslash G .
\ee

In order to interpret a point in $P^G_\Gamma$, i.e. a set of holonomies and fluxes on a graph which satisfy the discrete Gauss law, as a twisted geometry, one recognizes that the discrete Gauss law provides the closure relation (\ref{closure}) required for the Minkowski theorem. This implies that each vertex of a graph is represented geometrically as a polyhedron-shaped cell. Each flux $\X_e$ defines a face $f$ as described above, where the magnitude $X_e$ provides the area of $f$ and the direction $\hat{\X}_e$ defines the orientation of $f$. Notice the duality between vertices in the graph and cells of the twisted geometry, and likewise for edges of the graph and faces of the twisted geometry. The holonomies provide a notion of parallel transport between the twisted geometry cells, and tell us about the curvature of the overall three-geometry. One can find a map \cite{LSpeziale1} from the holonomy-flux data to a set of geometrical variables which describe the twisted geometry, and one can find a Poisson algebra for the new variables in order to define a twisted geometry phase space $T_\Gamma$. Now, for any given set of holonomies and fluxes on a graph, there is a unique set of polyhedra associated to the vertices, and the reverse statement also holds. This implies isomorphism between the holonomy-flux phase space and a twisted geometry phase space:
\be
P_\Gamma^G \cong T_\Gamma .
\ee

Let us now turn to the relationship between the holonomy-flux phase space and spinning geometries. A continuous phase space for general relativity is given in terms of the Ashtekar variables $(\A, \E) \in \mathcal{P}$ which define a spatial geometry embedded within spacetime. The one-form $\A \in \su(2)$ is a connection defining the extrinsic curvature of the spatial slice, and the two-form $\E \in \su(2)$ is the `electric' field\footnote{The electric field is closely related to the triad $\bm{e}$ on a spatial hypersurface, and can be written as $\E = [\bm{e}, \bm{e}]$.} which defines the intrinsic spatial geometry. The fields $(\A, \E)$ are subject to a continuous Gauss law $\mathcal{G}$ which generates $\SU(2)$-gauge transformations. By symplectic reduction we obtain an $\SU(2)$-gauge invariant phase space $\mathcal{P}^\mathcal{G} \equiv \mathcal{P} \sslash \mathcal{G}$.

In order to relate points in $P^G_\Gamma$ with a continuous geometry, we embed a graph within a regular three-dimensional cellular space $\Delta$ such that each vertex $v \in \Gamma$ lies within a cell $c$, and each edge $e \in \Gamma$ intersects a single face $f$. This gives the same duality between the graph and the cellular space as we had for twisted geometries. The key step now is to impose that the curvature of the connection vanishes $\bm{F}(\A)=0$ \textit{within} each cell $c$, which implies that the curvature is non-zero only on the one-skeleton $\Gamma^*$ of $\Delta$. Now, this flatness constraint generates its own gauge transformations on the fields, and we can again apply a symplectic reduction. This yields a finite-dimensional, truncated phase space $\mathcal{P}_{\Gamma^*}^\mathcal{G} \equiv \mathcal{P}^{\mathcal{G}} \sslash F_{\Gamma^*}$, where $F_{\Gamma^*}$ denotes the constraint which allows curvature only $\Gamma^*$, as well as the gauge transformations generated by this constraint. It is this truncated phase space which is isomorphic to the holonomy-flux phase space:
\be
P_\Gamma^G \cong \mathcal{P}^\mathcal{G}_{\Gamma^*} .
\ee 
What this tells us is that each point in $P_\Gamma^G$ has a one-to-one correspondence with an equivalence class of continuous geometies, and the equivalence class is defined by identifying field configurations that are related by gauge transformations:
\be
(\A, \E) \sim \left(g \A g^{-1} + g \rd g^{-1}, g (\E + \rd_A \bm{\phi}) g^{-1} \right).
\ee
Here $g \in \SU(2)$ is parametrizing a gauge transformation generated by the Gauss constraint, while the one-form $\bm{\phi}(x) \in \su(2)$ satisfying $\bm{\phi}(x) = 0$ for all $x \in \Gamma^*$ parametrizes a transformation generated by the flatness constraint. Notice that there is a rather large class of electric fields, i.e. intrinisic geometries, identified as the same point in the truncated phase space. Within each equivalence class is a spinning geometry which can be selected by asking that the three-geometry is piecewise-torsionless. That is we demand that $\rd_{A} \bm{e} =0$ within each cell, where $\bm{e}$ is the frame field. In this way we have a geometrical representation in terms of continuous fields $(\A, \E) \in \mathcal{P}^\mathcal{G}_{\Gamma^*}$ which are piecewise-flat and piecewise-torsionless, i.e. a spinning geometry\footnote{The fact that the spinning geometry fields $(\A, \E)$ are flat $\bm{F}(\A) = 0$ and torsionless $\rd_A \bm{e} = 0$ within each cell (where $\bm{e}$ is the triad), implies that we can write the geometry of each cell in terms of a group-valued field $a \in \SU(2)$ and an algebra-value field $\z \in \su(2)$ as follows:
\be
\A = a \rd a^{-1}, \qquad \qquad \E = a [\rd \z , \rd \z] a^{-1}.
\ee
This gives rise to a description of the local cell geometry in terms of the coordinate function $\z$, as used to define a spinning geometry above.}. Since we can always select the spinning geometry from an equivalence class in a well-defined manner, we in fact have a one-to-one correspondence between spinning geometries and the holonomy flux phase space $P^G_\Gamma$.

In this section we have described isomorphisms which relate the holonomy-flux phase space $P_\Gamma$ to both a twisted geometry and an equivalence class of continuous geometries. This implies that the truncated, continuous phase space is isomorphic to a twisted geometry phase space:
\be
\mathcal{P}^\mathcal{G}_{\Gamma^*} \cong T_\Gamma .
\ee
Now, since each point in the truncated space $\mathcal{P}^\mathcal{G}_{\Gamma^*}$ can be represented by a spinning geometry, we can extend this isomorphism to a one-to-one correspondence between a spinning geometry and a twisted geometry. One can visualize this as follows. Let us take a spinning geometry, and decompose it cell by cell. The data we have for each cell defines a polyhedron by the Minkowski theorem, and if we deform each of these spinning geometry cells into its associated polyhedron, we obtain a twisted geometry. Turning this argument around, one can view a spinning geometry as a way to form a continuous three-geometry from a twisted geometry.

We have then two isomorphic interpretations of the loop gravity phase space which possess different attributes of a Regge geometry. On the one hand, the spinning geometry is a continuous three-geometry which can represent a spatial hypersurface of spacetime. However, the cell boundaries take arbitrary curved shapes rather than the neat form of polyhedra. On the other hand, cells of the twisted geometry interpretation do take the form of polyhedra, but we lose the ability to describe a continuous three-geometry.

\section{Angular momentum}
\label{mom}
In loop gravity, fluxes measure the areas of surfaces and play an important role in determining local geometry. In the quantum theory, length \cite{length}, area \cite{area} and volume \cite{volume} operators are constructed from flux operators. In twisted geometries, the flux variables completely determine the shape of cells. In spinning geometries, fluxes only partially fix the cell shapes. The precise role that fluxes play in fixing a spinning geometry will be more clear at the end of the next section.

In addition to the relationship between fluxes and the areas of surfaces, these parameters are also closely related to angular momenta. This is because they satisfy the angular momentum Poisson algebra:
\be
\left\{ \X_f^i, \X_f^j \right\} = {\epsilon^{ij}}_k \X_f^k,
\ee
In this section we uncover another way in which flux is related to angular momentum.

Recall that each cell within a spinning geometry is assigned a set of flat coordinates $\z^c \in \su(2)$.
%These coordinates are related to the coordinates in a neighbouring cell $c'$ by:
%\be
%\label{z12}
%\bm{z}^{c^\prime}(s_{c'c}(x)) = h_{{c c^\prime}}^{-1} (\bm{z}^{c}(x)+\bm{a}_{c c^\prime}) h_{{c c^\prime}}, \hspace{0.5in} \forall x \in f_{c c^\prime},
%\,\, s_{c'c}(x)\in f_{c'c}.
%\ee
%where $h_{c c^\prime} \in \SU(2)$ is a group element associated to the face $f_{c c^\prime}$, and $\bm{a}_{c c^\prime} \in \su(2)$ is a algebra element corresponding to a translation.
In terms of these coordinate functions, the flux is given by \cite{FGZ}:
\be
\label{flux}
\bm{X}_{{c c^\prime}} = \frac{1}{2} \int_{f_{c c^\prime}} [\rd \bm{z}^{c} , \rd \bm{z}^{c}],
\ee
where the direction of integration is clockwise as seen from inside the cell $c$, and the bracket on the right hand side implies taking both the wedge product and $\su(2)$ commutator between elements\footnote{In coordinates this means
${X}_{{c c^\prime}}^{i} = \frac{1}{2} \epsilon^{ijk}\int_{f_{c c^\prime}} \rd \bm{z}^{c}_{j} \wedge \rd \bm{z}^{c}_{k}$}.

Recall that each face possesses two orientations so that the flux appears to have a different orientation from each of the two cells which share the face. The flux $\X_{{c^\prime c}}$ determined by $\bm{z}^{c^\prime}$ is given by:
\be
\label{X12}
\X_{{c^\prime c}} = -\frac{1}{2} \int_{f_{c^\prime c}} [\rd \bm{z}^{c^\prime} , \rd \bm{z}^{c^\prime}] = -h_{{c c^\prime}}^{-1} \X_{{c c^\prime}} h_{{c c^\prime}},
\ee
in agreement with (\ref{gluing}) above. The negative sign comes from reversing the direction of integration.

In the case when the links bounding the face are straight, $\bm{z}^c$ is linear along each link and $\bm{\xi}_{\l}^c\equiv \dot{z}^{c}(x)$ for all $x \in \ell$ is a constant vector associated to each link in the boundary of $c$.
For a closed face these vectors span a plane, the face is flat and we are in the Regge case.
In general this is not true, but we can still assign a vector to each link in the boundary of $c$. This follows from a very simple but extremely important remark:
If we integrate by parts in (\ref{flux}) we can write the flux as:
\be
\label{flux2}
\X_{{c c^\prime}} = \sum_{\l \in \partial f_{c c^\prime}} \J_{\l}^c, \qquad \qquad \J_{\l}^c := \frac{1}{2} \int_\l [\z^c, \rd \z^c],
\ee
which shows that we get a contribution $\J_\l^c$ from each link $\l_{c} \in \partial f_{cc'}$.
This means that we can in fact decompose the flux in terms of a sum of contributions associated with each edge
of the face.

Let us parameterize a link $\l(s)$ by the proper length $s$ to write:
\be
\label{linkmom}
\J_\l^c = \frac{1}{2} \int_{\l_{c}} \rd s [\z^c, \dot{\z}^c],
\ee
where $\dot{\z}^c \equiv \partial_s \z^c$. If we interpret $s$ as a time coordinate, then the integrand is a cross product between position and velocity, i.e. an angular momentum. Each $\J^c_\l$ is the angular momentum of a point particle in a 3d Riemannian `spacetime', integrated over the `worldline' given by the link $\l$. We shall refer to each $\J^c_\l$ as a \textit{link momentum}. In light of this, the flux associated to a face is given by the sum of the link momenta around its boundary.

In order to maintain agreement with (\ref{gluing}), the momentum associated to a link as seen in cell $c'$ is related to the momentum seen in cell $c$ via:
\be
\label{J12}
\J_{\l}^{c^\prime} = -h_{{c c^\prime}}^{-1} (\J_{\l}^c+[\bm{a}_{c c^\prime}, \D_{\l}^{c}]) h_{{c c^\prime}}.
\ee
where $\D_{\l}^{c} \equiv \z^c_{t(\l)} - \z^c_{s(\l)}$ is the difference between the coordinate function evaluated at the terminal $t(\l)$ and starting $s(\l)$ endpoints of the link.

As with the fluxes, link momenta are related to the oriented area of a two surface. When a link is straight one can perform the integral in (\ref{linkmom}) to obtain:
\be
\label{ReggeJ}
\J_{\l}^{c} = \frac{1}{2} [\z^c_{s(\l)} , \D_{\l}^{c}].
\ee
In this case we see that $\bm{J}_{\l}^{c}$ is the oriented area of a two-surface bounded by the link $\l$ and two straight edges joining the endpoints of $\l$ with the origin of the coordinate function $\bm{z}^c_\l$. Since this is the link momentum when the edge is straight, this is the value one obtains for a link in a Regge geometry.

If the link is not straight it is no longer true that we can express simply $\J_{\l}^{c}$ as a cross product with $\D_{\l}^{c}$.
If one thinks of $\D_{\l}^{c}$ as a total displacement vector, we see that the Regge contribution is analogous to the expression for the orbital angular momenta.
If one pushes this analogy further, it is natural to interpret the case where the link momenta is not Regge as being similar to the case where the total angular momenta contains a spin contribution. That is, in general we can decompose $\J$ into
an orbital or Regge component $\L$, and a spin contribution $\S$:
\be
\J^{c}_{\l} = \L^{c}_{\l} + \S^{c}_{\l},\qquad \L^{c}_{\l} = \frac{1}{2} [\z^c_{s(\l)} , \D_{\l}^{c}].
\ee
The spin contribution measure the deviation from Regge and vanishes when the link is straight.
We will see that the spin contribution literally corresponds to a spinning trajectory.

The decomposition of the fluxes $\X_{{c c^\prime}}$ in terms of link angular momenta  correspond to a general solution of the Gauss relation. We can interpret the Gauss law $\sum_{c'} \X_{{c c^\prime}} = 0$ as a discrete differential identity $ (\delta \X)_{c}=0$ whose solution is given by $ \X_{cc'}= (\delta \J)_{cc'}$ or $\X_{{c c^\prime}} = \sum_{\l \in \partial f_{c c^\prime}} \J_{\l}^c$. This solution follows from the fact that the discrete operation $\delta$ acts as a differential, i.e. $\delta ^{2}=0$.

This solution is however a local solution valid around every cell. Moreover, a priori the number of faces differs from the number of edges, so it is not obvious that we can always trade face fluxes for link momenta.
For instance in a triangulation we generically have more edges than faces. 
 Let us do an analysis to help understand this better. If we take the three-geometry to be homeomorphic to $S^3$, the number of links $\l^\#$, nodes $n^\#$, faces $f^\#$ and cells $c^\#$ satisfy the relation:
\be
c^\# - f^\#+\l^\# - n^\# = 0.
\ee
Recall that the fluxes satisfy a closure relation (\ref{closure}). This constraint is applied in each cell except one, since the last one is redundant. Therefore the total number of degrees of freedom in the fluxes of a spinning geometry is $X^\# = 3 \times (f^\# - c^\# +1)$.

There are $J^\# = 3 \times \l^\#$ degrees of freedom in a set of link momenta. The counting of extra degrees of freedom in link momentum versus fluxes is $J^\# - X^\# = 3 \times(n^\# - 1)$. These extra degrees of freedom are accounted for in the kernel of the map (\ref{flux2}) which is invariant when each of the link momenta undergoes the transformation:
\be
\label{Jtrans1}
\bm{J}_{\l}^c \longrightarrow \bm{J}_{\l}^c + \bm{\zeta}_{t(\l)}^c - \bm{\zeta}_{s(\l)}^c,
\ee
where $\bm{\zeta}_{t(\l)}^{c}$ is a vector at the terminal node of the link and $\bm{\zeta}_{s(\l)}^{c}$ is a vector at the starting node of the link. Over the entire cellular decomposition there is one such independent vector at each node, which accounts for the remaining $3 \times(n^\# - 1)$ degrees of freedom. The $-1$ here comes from the fact that one of these shifts can be accounted for by an overall translation.
This analysis shows that there are as may degrees of freedom in a set of fluxes subject to the Gauss constraints as there are in the link momenta, modulo an invariance related to the translation of nodes.
Indeed, The transformation in (\ref{Jtrans1}) is defining an equivalence class of link momenta, where each member of the class maps to the same set of flux data. If we specify a particular set of link momenta we are choosing one member of this equivalence class.

Note finally that the kernel of the map in (\ref{flux2}) is related to shifts in the coordinate functions. Looking at (\ref{linkmom}), if we translate the coordinates $\bm{z}^c \longrightarrow \bm{z}^c + \bm{a}$ the link momenta transform as:
\be
\label{Jtrans2}
\J_{\l}^{c} \longrightarrow \J_{\l}^{c} + \frac{1}{2} [\bm{a} , \z^c_{t(\l)}] - \frac{1}{2} [\bm{a} , \z^c_{s(\l)}].
\ee
This implies that the vectors transforming the link momenta in (\ref{Jtrans1}) are given at each node $n$ by $\bm{\zeta}_{n}^c = \frac{1}{2} [\bm{a} , \z^c_{n}]$.

\section{Minimizing link lengths}
\label{min}
As shown in \cite{FGZ} any piecewise flat geometry, is isomorphic to an equivalence class of holonomy-flux data which represents a single point in the $\SU(2)$-gauge invariant loop gravity phase space $P_{\Gamma}$. 
From this perspective it is still possible to  to deform the geometry of each individual cell without altering the link momenta. Notice that if we make deformations which leave the link momenta unchanged, then the fluxes remain fixed as well. Any two geometries related by such a deformation correspond to the same point in $P_{\Gamma}$. Our main goal is to explore this freedom and study the extent to which this ambiguity can be fixed by extra geometrical requirements. Is there a way to unambiguously fix the cell shapes to select a unique geometry to represent a point in $P_{\Gamma}$?

We also would like to understand how the holonomies enter the geometrical picture.
A priori, these holonomies are  unchanged by deformations of cell boundaries \cite{FGZ}. The holonomy associated to a face $f_{cc'}$ represents the parallel transport from the interior of a cell $c$ to the interior of a neighbouring cell $c'$. In a spinning geometry, the connection is flat on cells and faces, so the holonomies depend only upon data at their endpoints. But as we will see the holonmies constrain the shape of each internal edge.
Before analyzing this point we first focus on how fluxes depend upon the face shapes.

The Gauss law of loop gravity implies that the flux associated to a face depends only the links on the boundary, i.e. if one fixes the links on the boundary, any choice of face bounded by these links will yield the same flux \cite{FGZ}. This is shown explicitly in the expression of the flux given in (\ref{flux2}). What this means is that we need only consider deformations of links in the one-skeleton, and can ignore what is happening to the faces as long as we do not examine the continuity of the normal component of the frame field across faces.

Now, the one-skeleton of a Regge geometry is composed of straight links, so that each link is the shortest path between the nodes at its endpoints. Let us use this idea to try to reduce the ambiguity in the choice of one-skeleton for a spinning geometry by minimizing the length of each link while keeping the nodes fixed. At the same time, let us impose a constraint which keeps the link momenta fixed so that we maintain a correspondence with the same point in the phase space $P_{\Gamma}$.

This can be achieved by introducing the following action which sums over all of the link lengths and includes a constraint to fix each link momentum:
\be
\label{total action}
S=\sum_\l S_\l, \qquad\quad   S_\l = \int_\l |\dot{\bm{z}}_\l^{c}|\rd s +\bm{\omega}_\l^{c} \cdot  \left( \bm{J}_\l^{c} - \frac{1}{2}\int_\l (\z_\l^{c}\times \dot{\z}_\l^{c}) \rd s\right),
\ee
where $\bm{\omega}_\l^{c}\in \su(2)$ is a Lagrange multiplier implementing the constraint which fixes $\bm{J}_{\l}^{c}$. Notice there is only one term per link although each link $\l$ is attached to several cells. This is why we introduce a subscript on the coordinate functions $\z_\l^{c}$ to denote a choice of which coordinates are to be used for each term. Using the relations (\ref{z12}, \ref{J12}), this action does not depend upon this choice, so long as we identify that the Lagrange multiplier defined in one cell is related to that of a neighbouring cell by:
\be
\w^{c'}_\l = -h_{cc'}^{-1} \w^{c}_\l h_{cc'}.
\ee

Varying the action we find:
\be
\label{variation}
\delta S =\sum_{\l} \int_\l  \delta{\z}_\l^c \cdot\left( \dot{\bm{v}}^c_{\l}  - \bm{\omega}^c_{\l}\times \dot{\z}^c_{\l}\right) \rd s
\ee
where $\bm{v}^c_{\l} \equiv \frac{\dot{\z}^c_{\l}}{|\dot{\z}^c_{\l}|}$ is the `proper velocity', treating $s$ as the proper time and drawing upon our analogy with 3d point particles. Setting this variation to zero provides an equation of motion for each link:
\be
 \dot{\bm{v}}^c_{\l} = \bm{\omega}^c_{\l}\times \dot{\z}^c_{\l}.
\ee
Let us now study the equation for a single link.

\subsection{Analysis of a single link}
Since we are analyzing a single link for some choice of coordinate function $\z_\l^c$, we drop the super- and sub-scripts $c$ and $\l$ for simplicity. Let us first assume that $\omega \neq 0$. If this is not satisfied the equation of motion simply tells us that the link has to be a straight line and we are back to the Regge case. When $\omega \neq 0$ it will be convenient for us to chose a parametrization of the curve $s=\phi$ where $\phi$ is $\omega$ times the proper length, that is:
\be
\label{normalization}
\dot{z}_\phi = \omega^{-1}.
\ee

In this parametrization the equation of motion is simply:
\be
\label{eom}
\ddot\z_\phi = \bm{\hat\omega} \times \dot\z_\phi.
\ee
This equation can be easily solved. First of all, it implies that we have the following three conserved quantities:
\be
K\equiv \bm{\hat\omega}\cdot \dot\z_\phi,
\qquad \qquad
\omega \equiv \dot{z}_\phi^{-1},
\qquad \qquad
r \equiv \ddot z_\phi.
\ee
Moreover $\bm{r}_\phi \equiv -\ddot\z_\phi$ is a vector orthogonal to $\hat{\bm{\omega}}$ and a solution of the equation $ \dot{\bm{r}}_{\phi} = \bm{\hat\omega} \times {\bm{r}}_{\phi}$. Since this implies that $\ddot{\bm{r}}_{\phi} = \bm{\hat\omega} \times {\dot{\r}}_{\phi} = -{\bm{r}}_{\phi} $, we have:
\be\label{rphi}
{\r}_\phi = \cos\phi\, \bm{r}_{0} + \sin\phi\, (\bm{\hat\omega} \times {\bm{r}}_{0}).
\ee
The solution to (\ref{eom}) is therefore given by:
\be
\label{sol}
\z_\phi = \bm{c} + K \phi \bm{\hat\omega} +  \bm{r}_{\phi},
\ee
where $\bm{c}\in \su(2)$ is a constant. This is the parametric equation for a helix.

\begin{figure}[htb!]
\begin{center}
\includegraphics[width=0.25\linewidth]{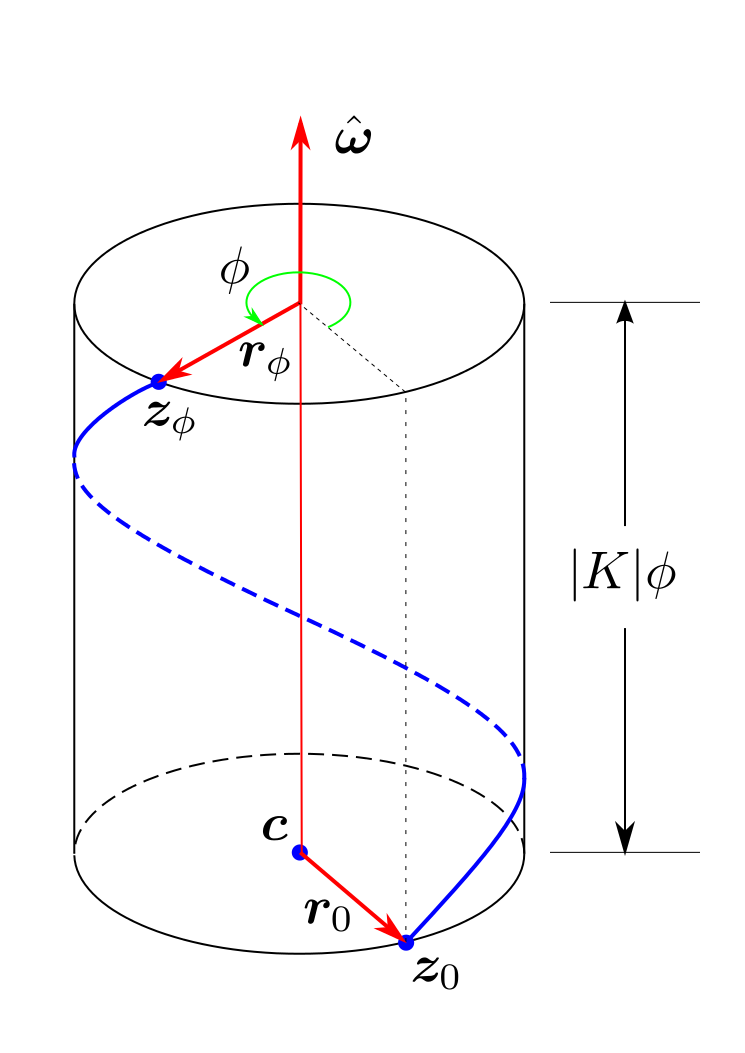}
\caption{A helix (in blue) wrapping around a cylinder.}
\label{cylinder}
\end{center}
\end{figure}

In order to  visualize the situation and understand the meaning of the different parameters,
let us recall that a helix can be drawn on the boundary of a cylinder, wrapping around it.
This cylinder possesses an axis given by the direction $\hat{\bm\omega}$, and a radius given by $r$. The parameter
$\phi$ is the angle, counterclockwise about $\hat{\w}$, elapsed from the initial point $\z_0$ to the point $\z_\phi$. The radial vector $\bm{r}_{\phi}$ is the position of the point on a circle perpendicular to the axis of the cylinder.
The parameter $K$ is related to the height $H$ of the cylinder by
$H =|K \phi|$.
Finally, the point $\bm{c}$ is on the axis at the `bottom' of the cylinder. See fig. (\ref{cylinder}) for an illustration.

The normalization condition (\ref{normalization}) implies a relation between $\omega$, $K$ and $r$:
\be
1= (K\omega)^{2} + (r\omega)^{2}.
\ee
$K \omega$ represents the linear velocity along the cylinder axis and $r \omega$ the angular velocity.

There is a redundancy in allowing $\phi$ and $K$ to take either sign. First of all, notice from (\ref{sol}) that $K$ is positive when the axial component of the velocity is directed along $\hat{\w}$, and negative when it runs opposite to $\hat{\w}$. Now, the helix generated by a positive angle $\phi$ is different than the one generated by a negative angle. However, since $\phi$ is defined to be counterclockwise about $\hat{\w}$, the transformation $\phi \rightarrow -\phi$ is equivalent to  $\w \rightarrow -\w$ with $K \rightarrow -K$. We eliminate this redundancy by taking $\phi>0$ while allowing the axis $\hat{\w}$ to point in any direction.

We can express the displacement vector $\D_\phi \equiv \z_\phi - \z_0$ which is connecting the start of the helix to the point $\z_\phi$ along a straight line in terms of the helix parameters:
\be
\D_\phi = K \phi \hat{\w} + \bm{r}_{\phi}-\bm{r}_{0}.
\ee
This allows us to express the position and velocity as:
\be
\label{zsol}
\z_\phi = \z_0 + \D_\phi \qquad \qquad \dot{\z}_\phi = \dot{\D}_\phi.
\ee
These equations allow for a simple calculation of the link momentum in terms of a minimal set of helix parameters.

\subsection{Helix parameters}
We are now in a position to give the parameters of the helix. In order to do this, we first develop an expression for the link momentum in terms of the quantities introduced above
.
Recall that the link momentum is given by:
$
\bm{J} =\frac12 \int_l  (\bm{z} \times \dot{\z}) \rd \phi.
$
Using $\Phi>0$ to denote the total angle elapsed from the start to the end of the helix, i.e. $\phi \in [0,\Phi]$, a direct evaluation gives:
\be
\J &=&\frac12 \int_0^{\Phi} \rd \phi \left( \z_0 \times \dot{\D}_\phi + \D_\phi \times \dot{\D}_\phi \right) \nonumber \\
&=& \frac12 \int_0^{\Phi} \rd \phi \left(\z_0 \times \dot{\D}_\phi + K \phi \ddot{\r}_\phi - K \dot{r}_\phi + K \hat{\w} \times \r_0 + (\r_\phi - \r_0) \times \dot{\r}_\phi \right) \nonumber \\
&=& \frac12 \int_0^{\Phi} \rd \phi \left( \z_0 \times \dot{\D}_\phi + \partial_\phi (K \phi \dot{\r}_\phi) + K(\hat{\w}\times \r_0- 2\dot{\r}_\phi) + r^2 \hat{\w} - \r_0 \times \dot{\r}_\phi \right).
\ee
Now, performing the integral and using the above expression (\ref{rphi}) for $\r_\Phi$, we obtain:
\be
\J &=& \frac12 \z_0 \times \D_\Phi + \frac12 \left( K \Phi \dot{\r}_\Phi + K \Phi \hat{\w}\times \r_0 - 2 K( \r_\Phi - \r_0) + r^2 \Phi \hat{\w} - \r_0 \times \r_\Phi \right), \nonumber\\
&=& \frac12 \z_0 \times \D_\Phi + \frac12 r^2 \left( \Phi - \sin \Phi \right) \s_0 \nonumber \\
&&+ r K \left( 1 - \cos \Phi - \frac{\Phi}{2} \sin \Phi \right)\s_1 + r K \left( \frac{\Phi}{2} + \frac{\Phi}{2} \cos \Phi - \sin \Phi \right) \s_2 ,
\ee
where we have introduced a shorthand notation for the orthonormal `helix' basis:
$
\s_i \equiv \left( \hat{\w}, \hat{\r}_0, \hat{\w} \times \hat{\r}_0 \right).
$

With the help of some trigonometric identities we can write the above expression in a more simple form:
\be
\label{Jsol}
\J &=& \L + \S,\qquad
\label{sRsol}
\L =  \frac12 \z_0 \times \D_\Phi,\\
\label{sJsol}
\S &=& r^2 f_\varphi \s_0 + 2 r K \varphi g_\varphi \s_{\varphi},
\ee
where $\varphi \equiv \Phi / 2$ is half of the total angle, $ \s_{\varphi}\equiv \left( - \sin \varphi \s_1+ \cos \varphi \s_2 \right)$ and we have defined two functions of this angle:
\be
f_\varphi \equiv \varphi - \cos \varphi \sin \varphi, \qquad \qquad g_\varphi \equiv \cos \varphi - \frac{\sin \varphi}{\varphi}.
\ee
The Regge contribution $\L$ is the link momentum one would obtain for a straight link as in (\ref{ReggeJ}), and
the non-Regge contribution $\S$ is giving the deviation from this value. The non-Regge contribution is invariant under translation $\bm{z} \to \bm{z} + \bm{a}$ while the Regge contribution is not.

Note that we can express the displacement vector in the helix basis:
\be
\label{Dsol}
\D_\Phi \equiv 2K\varphi \s_0 + 2r \sin\varphi\s_{\varphi}.
\ee
From the above equations (\ref{Jsol}--\ref{Dsol}), it is apparent that a minimal set of parameters for determining $\J$ is given by $(\z_0, r, K, \varphi)$ and the helix basis $\s_i$, which is given by a pair of orthogonal unit vectors $(\hat{\w},\hat{\r}_{0})$.

Following Penrose and Rindler \cite{Penrose} we call the pairs $(\hat{\w},\hat{\r}_{0})$ a {\it flag}, where $\hat{\w}$ is the pole of the flag and $\hat{\r}_{0}$ the direction of the flag.
Using the Hopf fibration, a flag is equivalent to a point in $S^{3} / \mathbb{Z}_{2}$ where the $\mathbb{Z}_{2}$ action is the inversion. This can be seen as follows. We can label a point in $S^{3}$ by a pair of complex numbers $|x \rangle =(x_{0},x_{1})$ satisfying the normalization condition $ |x_{0}|^{2} + |x_{1}|^{2}=1$.
We can label a point in $S^{2}$ by a vector $| y )=(y_{0}, y = y_{1}+iy_{2}) \in S^2$ where $y_0 \in \mathbb{R}$ and $y \in \mathbb{C}$ satisfy the condition $y_{0}^{2} + |y|^{2}=1$.
The Hopf fibration is a many-to-one map which sends circles in $S^3$ to points in $S^2$.
It is given by $\pi(|x\rangle)=( |x_{0}|^{2} - |x_{1}|^{2}, 2 x_{0}\bar{x}_{1})$.
We can extend this map to a map from $S^{3}$ to a flag $(\hat{\w},\hat{r}_{0})$. The kernel of this map is the $\mathbb{Z}_{2}$ inversion and it is given by:
\be
\hat{\w} = ( |x_{0}|^{2} - |x_{1}|^{2}, 2 x_{0}\bar{x}_{1}),\qquad
\hat \r_{0} =(x_{0}x_{1}+ \bar{x}_{0}\bar{x}_{1}, \bar{x}_{1}^{2}-x_{0}^{2}).
\ee
It can be checked explicitly that $\hat{\w}^{2} =\hat{\r}_{0}^{2}=1$ and $\hat{\r}_{0}\cdot \hat{\w}=0$ as required.
The knowledge of a flag $(\hat{\w},\hat{\r}_{0}) $ determines $ (x_{0},x_{1})$ only up to a global sign, accounting for the $\mathbb{Z}_{2}$ symmetry noted above. Notice that there are three degrees of freedom in a flag.

\begin{figure}[htb!]
\begin{center}
\includegraphics[width=0.4\linewidth]{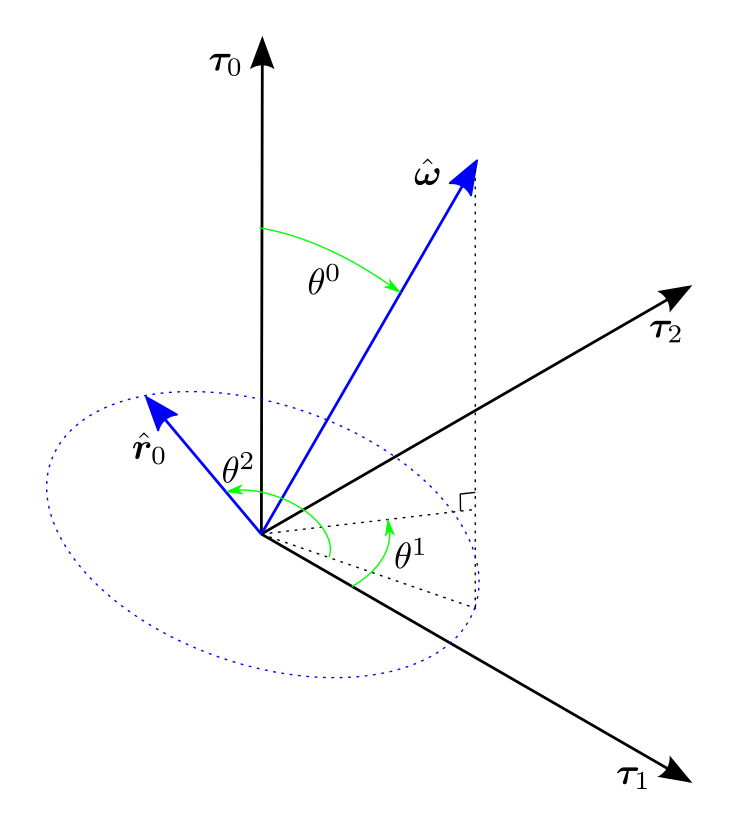}
\caption{The flag $(\hat{\w}, \hat{\r}_0$) (in blue) with respect to a fixed basis $\t_i$. To measure $\theta^2$, one projects $\hat{\w}$ down the direction of $\t_0$ into a plane perpendicular to $\hat{\w}$. $\theta^2$ is the angle from this projection to $\hat{\r}_0$, measured within the plane perpendicular to $\hat{\w}$.}
\label{flag}
\end{center}
\end{figure}

We can also write a flag in term of three angles $\theta^i$ that are relative to a fixed basis $\t_i$ as shown in fig. \ref{flag}. The first of these is the angle $0\le \theta^0 < \pi$ between $\hat{\w}$ and $\t_0$; the second $0\le \theta^1 < 2\pi$ is the angle between $\t_1$ and the projection of $\hat{\w}$ onto the $(\t_1, \t_2)$ plane; the third angle $0\le \theta^2 < 2\pi$ gives the orientation of $\hat{\r}_0$ in the plane perpendicular to $\hat{\w}$. With these angles the flag can be written as:
\be
\hat{\w} &=& \cos \theta^0 \t_0 + \sin \theta^0 \cos \theta^1 \t_1 + \sin \theta^0 \sin \theta^1 \t_2; \\
\hat{\r}_0 &=& \sin \theta^0 \cos \theta^2 \t_0 + (\sin \theta^1 \sin \theta^2 - \cos \theta^0 \cos \theta^1 \cos \theta^2 )\t_1 - (\cos \theta^0 \sin \theta^1 \cos \theta^2 + \cos \theta^1 \sin \theta^2) \t_2 .\nonumber
\ee
These angles are related to the complex parameters above via:
\be
x_0 = \pm \left(\cos \frac{\theta^0}{2} \right) \exp \left(\frac{i}{2} (\theta^1+\theta^2) \right), \qquad x_1 =  \pm \left(\sin \frac{\theta^0}{2}\right) \exp \left(\frac{i}{2}(\theta^2-\theta^1) \right).
\ee

There is yet a third way to represent the flag data.
Notice that the angles $\theta^i$ are giving the helix basis $\s_i$ in terms of the fixed basis $\t_i$. The two bases are related by a (passive) rotation, i.e. $\t_i =  e_i{}^j \s_j$ where $e_i{}^j = \t_i \cdot \s^j$. The transpose $(e^T)_j{}^i= \t^i \cdot \s_j= e^{i}{}_{j}$ is the inverse rotation and satisfies $(e^T)_k{}^j e_j{}^i  = e_k{}^j (e^T)_j{}^i  = \delta^i_k$. Knowledge of this flag matrix allows one to determine the angles $\theta^i$, implying that $e_i{}^j$ is equivalent to the flag data.

The passive rotation of a basis is equivalent to an active inverse rotation of the vector components. Given the six helix parameters $(r, K, \varphi, \theta^i)$ one can determine the fixed-basis components $\D_\Phi \cdot \t^i$ and $\S \cdot \t^i$ using the flag matrix and equations (\ref{Jsol}--\ref{Dsol}):
\be
\label{6eqJ}
\S \cdot \t^i &=& e^i{}_j (\S \cdot \s^j), \\
\label{6eqD}
\D_\Phi \cdot \t^i &=& e^i{}_j (\D_\Phi \cdot \s^j).
\ee

We have shown that there are nine degrees of freedom $(\z_0, r, K, \varphi, \theta^i)$ in the link momentum $\J$. We call this data the helix parameters since they define a unique helix via equation (\ref{zsol}). Using the six equations (\ref{6eqJ}, \ref{6eqD}) along with (\ref{Jsol}--\ref{Dsol}), we can determine $\J$ from the helix parameters. In other words, these equations provide a map:
\be
(\z_0, r, K, \varphi, \theta^i) \rightarrow (\z_0, \D_\Phi, \S) ,
\ee
where the data on the right hand side determines the link momentum $\J$. But to what extent does knowledge of $\D_\Phi$ and $\S$ in a fixed basis allow us to determine the helix parameters? Given $(\z_0, \D_\Phi, \S)$, can we invert this map to find a corresponding helix?
This question is at the heart of what we hope to accomplish, and we address it in the next subsection.

\subsection{Determining the helix parameters}
Let us now take take $(\z_0, \D, \S)$ as given for a single link (dropping the subscript from $\D_\Phi$), and from this data determine the helix parameters.

First of all, we remark that if $\S=0$, then the link is straight (a trivial helix) and the link momentum is given entirely by the Regge contribution $\J = \L = \frac12 \z_0 \times \D$. The parameters $(r, K, \varphi)$ no longer play any role so that the data defining the link is simply the position of the nodes $(\z_0, \D)$. In this way the inverse mapping is trivial for a straight link.

\begin{figure}[htb!]
\begin{center}
\includegraphics[width=0.8\linewidth]{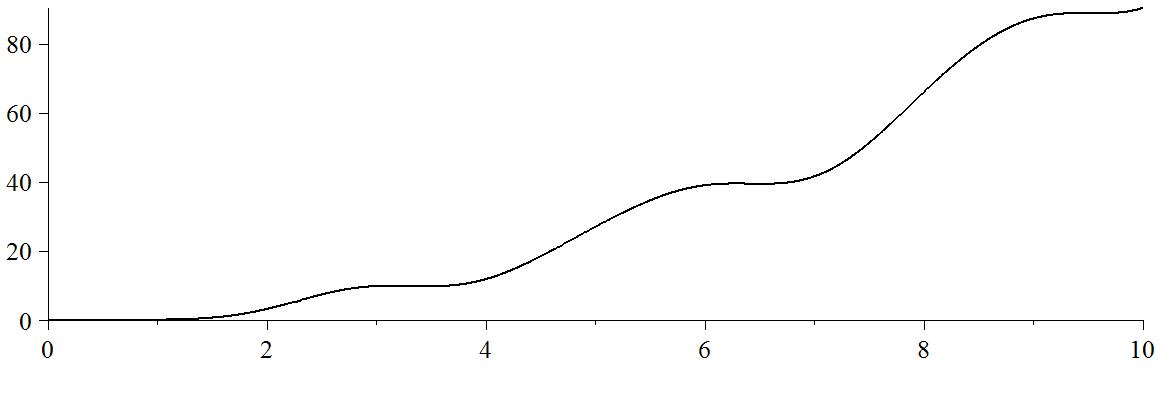}
\caption{A plot of $\Delta_{\varphi}=(f_\varphi^2 - 4 g_\varphi^2 \sin^2 \varphi)$, which is positive for all $\varphi>0$.}
\label{plot1}
\end{center}
\end{figure}

The interesting analysis is for $\S \ne 0$. In this case, we can choose a convenient fixed basis: $\t_0 = \hat{\D}, \t_1 = \widehat{\D \times \S}, \t_2 = \t_0 \times \t_1$. In this basis we have:
\be
\label{fixedJD}
\D \cdot \t^i &=& (D, 0, 0), \qquad
\S \cdot \t^i =  (S \cos \delta, 0, -S \sin\delta).
\ee
Here ${S}$ is the modulus of $\S$,  ${S} \cos\delta$ is the portion of the link momentum which is parallel to $\D$ while $S \sin\delta = |\hat{\D} \times {\S}| $ is the perpendicular component.
Squaring
%(\ref{6eqD}) and using (\ref{fixedD}) and
(\ref{sJsol}) and (\ref{Dsol}) gives the following equations for $K$ and $r$ in terms of $D,S$ and $\varphi$:
\be
D^{2}&=& 4K^{2}\varphi^{2} + 4 r^{2}\sin\varphi^{2}, \\
S^2 &=& f_\varphi^2 r^{4} + 4K^{2}\varphi^{2} r^{2}g_\varphi^2.
%\left(\frac{K}{D_\Phi}\right)^2 = \frac{1 - 4 \left( \frac{r}{D_\Phi} \right)^2 \sin^2 \varphi}{4 \varphi^2}.
\ee
Solving the first equation determines  $K$ only up to a sign\footnote{Explicitely $ K^{2}=\left(\frac{D^{2} - 4 r^2 \sin^2 \varphi}{4 \varphi^2}\right)$.} which we shall determine below.
Substituting the value for $K$ in the second equation we obtain  a quadratic equation for $r^2$. Only one of the solutions yields a real positive value. This gives us the expressions of $r$ and $K$ as functions of $\varphi$:
\be
\label{r2eq}
r^2_{\varphi} &=&D^{2}\left( \frac{-g_\varphi^2  + \sqrt{g_\varphi^4  +4 s^{2} \Delta_{\varphi}}}{2 \Delta_{\varphi}} \right),\\
K^2_{\varphi} &=&D^{2}\left( \frac{f_\varphi^2 - 2 \sin^2 \varphi\left(g_{\varphi}^{2}+  \sqrt{g_\varphi^4 +4 s^2 \Delta_{\varphi}}\right)}{4\varphi^{2} \Delta_{\varphi}} \right) \label{K2eq} ,
\ee
where we have introduced a dimensionless parameter $s\equiv {S}/D^{2}$ and the function $\Delta_{\varphi}\equiv (f_\varphi^2 - 4 g_\varphi^2 \sin^2 \varphi)$.
Note that since $(f_\varphi^2 - 4 g_\varphi^2 \sin^2 \varphi) >0$ for all $\varphi>0$ the right hand side is always real and well-defined. See fig. \ref{plot1} for a plot of this function.
Note that for small  $ s $ we have:
\be
r^2_{\varphi} \approx \frac{ {S}^{2}}{g_\varphi^2 D^{2}}, \qquad K^{2}\varphi^{2}\approx \frac{D^{2}}{4} .
\ee

This leaves us to solve for the angle $\varphi$ and the flag matrix.
In order to do so we introduce the following angles $(\alpha_{\varphi},\beta_{\varphi})$, solutions of\footnote{ More precisely we demand that
$  \cos \alpha_{\varphi} =  \frac{2 K_{\varphi} \varphi}{D}$, $ \sin\alpha = \frac{2 r_{\varphi} \sin\varphi }{D}$ and that
$\cos \beta_{\varphi} =  \frac{r^{2}_{\varphi}f_{\varphi}}{\mathcal{J}}$, $\sin\beta = \frac{2 r_{\varphi} K_{\varphi}\varphi g_{\varphi}}{\mathcal{J}}$.}:
\be
\tan\alpha = \left(\frac{ r_{\varphi}}{ K_{\varphi} \varphi}\right) \sin\varphi,\qquad \tan \beta = \left(\frac{ K_{\varphi}\varphi }{r_{\varphi}}\right)\frac{2 g_{\varphi}}{f_{\varphi}}.
\ee
These angles gives us the decomposition of the edge momentum and displacement vector in terms of the helix basis from (\ref{6eqJ}, \ref{6eqD}) along with (\ref{sJsol},\ref{Dsol}):
\be
\bm{D}= D (\cos\alpha \s_{0} + \sin \alpha \s_{\varphi}),\qquad
\S = {S} (\cos\beta \s_{0} + \sin \beta \s_{\varphi}).
\ee
Let us denote by $R(\t, \alpha)$ a rotation  about the $\t$ axis by an angle of $\alpha$.
It is defined by $\partial_{\alpha}R(\t, \alpha)(\s)= [\tau, R(\t, \alpha)(\s)]$.
We can  express the previous relations as:
\be
\bm{D}= D R(\s_{0},\varphi) R(\s_{1},\alpha)(\s_{0}),\qquad \S ={S} R(\s_{0},\varphi) R(\s_{1},\beta)(\s_{0}).
\ee
This shows that the rotation $R\equiv R(\s_{0},\varphi) R(\s_{1},\alpha)$ maps the helix basis onto the fixed basis (\ref{fixedJD}).

\begin{figure}[htb!]
\begin{center}
\includegraphics[width=0.8\linewidth]{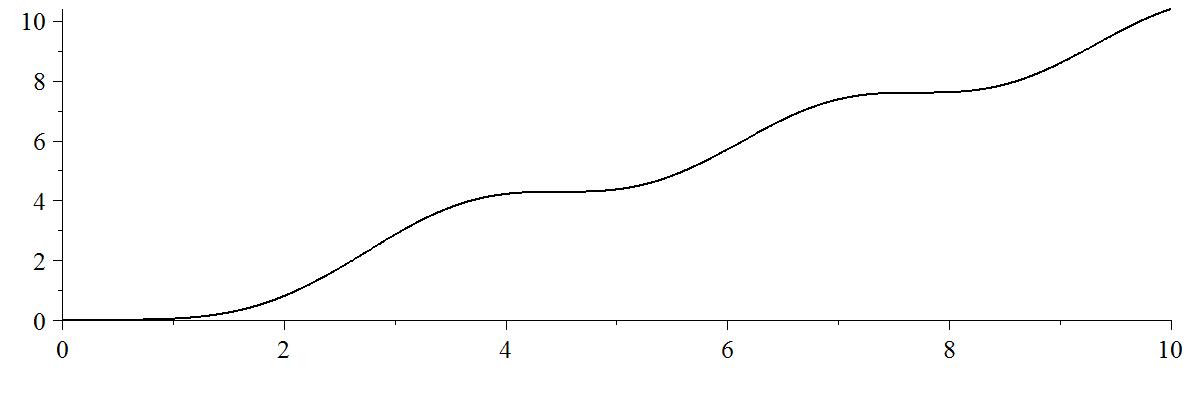}
\caption{A plot of $f_\varphi + 2 g_\varphi \sin \varphi$, which is positive for all $\varphi>0$.}
\label{plot2}
\end{center}
\end{figure}

Now that we have determined $(K,r)$ and the flag in terms of $\varphi$ and $(\bm{D},\S)$, the only task left is to determine $\varphi $ in terms of
$(\bm{D},\S)$. This follows from the equation $\delta = \alpha-\beta$ where  $\S\cdot {\bm{D}}=SD \cos\delta$ which gives us the final constraint:
\be\label{xieq}
\S\cdot {\bm{D}}= 2(r^{2}_{\varphi}K_{\varphi}\varphi)(f_{\varphi}+ 2 g_\varphi \sin\varphi).
\ee
This equation gives us $\varphi$ in terms of $(\D, \S)$. We have checked that $f_\varphi + 2 g_\varphi \sin \varphi>0$ for $\varphi >0$ (see fig. \ref{plot2} for a plot), which allows us to determine that the sign of $K$ is $\sgn(K) = \sgn(\S \cdot {\bm{D}})$.

\begin{figure}[htb!]
\begin{center}
\includegraphics[width=0.8\linewidth]{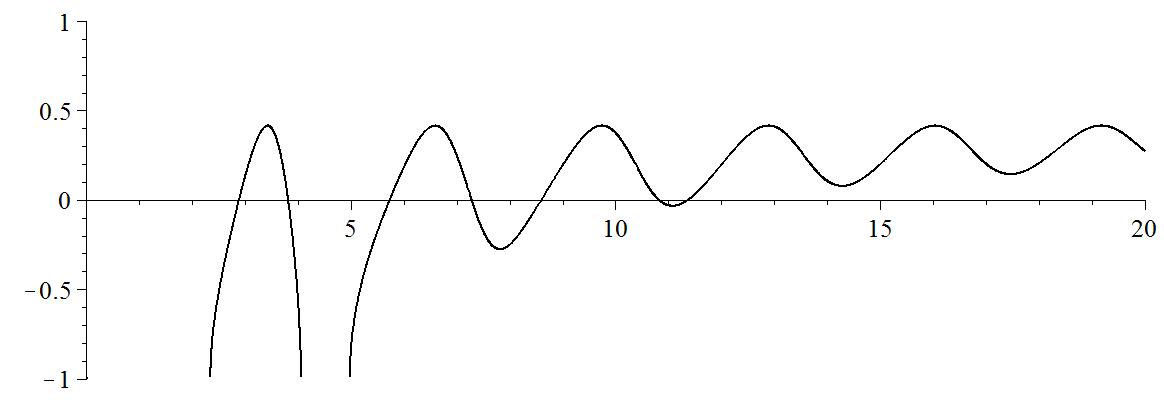}
\caption{A typical plot of $ 2(r^{2}_{\varphi}K_{\varphi}\varphi)(f_{\varphi}+ 2 g_\varphi \sin\varphi)- \S\cdot {\bm{D}}$ where $s\cos\delta = s\sin\delta = 1$. Notice there are seven solutions for $\varphi$.}
\label{plot}
\end{center}
\end{figure}

We checked numerically for solutions to this equation over a range of values for the parameters $-1000~\le~s~\cos\delta~\le~1000$ (excluding the case of $s \cos\delta = 0$ since this is inconsistent with $\varphi>0$ in (\ref{sJsol}), as required for a non-trivial helix) and $0~\le~s~\sin\delta~\le~1000$. We found in all cases that there is at least one intersection, and in general there actually many possible values of $\varphi$ which solve this equation. See fig. (\ref{plot}) for a typical plot of (\ref{xieq}). In minimizing the action, we have in fact found multiple local minima rather than a single global minimum.

This analysis shows that
 we can always invert the equations (\ref{6eqJ}, \ref{6eqD});
 given $(\z_0, \D, \S)$ there always exists a discrete set of  helix parameters $(r, K, \varphi, \theta^i)$.
 We can fix the ambiguity by always choosing the helix with the
 %minimal value of $\varphi$, i-e
 minimal length.
 This means that once the end points are given, there exists a one-to-one correspondance between a choice of link momentum and a choice of helix stretching between these end points. In other words, if we fix two nodes and specify a link momentum $\J$, we can always take the link to be in the form of a helix whose total angular momentum $\frac12\int (z\times \dot{z})$  is $\J$.

\subsection{Analysis over the entire one-skeleton}

Now that we have completed our analysis of a single link, let us extend this analysis over the entire one-skeleton of a spinning geometry.
The result of the previous section shows that one can deform each link of the one-skeleton into a helix without changing the link momenta. Recall that such deformations also leave the holonomies and fluxes unchanged, so that the resulting geometry where each link is now a helix corresponds to the same point in $P_{\Gamma}$. As argued in section \ref{flatdef}, these spinning geometries can be seen as a way to join the polyhedral cells of a twisted geometry into a continuous three-geometry, achieved by twisting the links of polyhedra into helices.

In this subsection we need to restore the link $\l$ and cell $c$ super- and sub-scripts. The quantities that were labeled $(\D,\J)$ above shall here be denoted $(\D_\l^{c},\J_{\l}^{c})$, where $\l$ denotes the link and $c$ denotes the reference frame attached to the cell $c$.

As we have, seen the loop gravity data consists of a set $(\bm{X}_{cc'},h_{cc'})\in \mathfrak{su}(2)\times \mathrm{SU}(2)$ for each pair of cells, modulo the relations
\be
h_{c'c}=h_{cc'}^{-1},\qquad \bm{X}_{c'c} = - h_{c'c}\bm{X}_{cc'}h_{cc'}.
\ee
The question is whether we can construct from this data a flat spinning geometry?
We have seen that given $(\D_{c}^{\l}, \J_{c}^{\l})$ compatible with the fluxes such that $\sum_{\l \in c}\J_{c}^{\l} = \bm{X}_{cc'}$,
  we can  construct for every edge $\l \subset c$ a helix that is compatible with this data.
  This means that we can construct a map $\bm{z}_{c} : c \to \mathbb{R}^{3}$ for every cell $c$.
  This map is such that the one-skeleton of its boundary consists of helices, while the shape of the faces are undetermined at this stage. In fact, there are many possible choices of boundary helices which are compatible with a given sets of fluxes.
  
Twisted geometries rely on the Minkowski theorem \cite{Minkowski}, which states that any set of fluxes satisfying the closure condition represents a unique polyhedron (up to translations). Because of this theorem, we know that it is always possible to choose only straight edges for a single cell. However this is no longer possible when we start to glue the cells together. The question is then, under what conditions on compatible $(\D_{c}^{\l}, \J_{c}^{\l})$ can we  consistently glue cells together, and what is the form of the gluing map $s_{cc'}$?

We have seen that consistency of gluing requires that
they satisfy the relations:
\be
\bm{z}^{c^\prime}(s_{cc'}(x) ) = h_{{c c^\prime}}^{-1} (\bm{z}^{c}(x)+\bm{a}_{c c^\prime}) h_{{c c^\prime}}, \hspace{0.5in} \forall x \in f_{c}.
\ee
Let us now consider a link $\l$ which is common to all the cells $c_{1},\cdots c_{n}$, that is we assume that $\l \equiv f_{{c}_{1}}\cap \cdots \cap f_{{c}_{n}}$ under the gluing maps.
This means that for $x\in \l_{c_1}$ we have a map $ S_{c_{1}c_{i}}(x)\equiv s_{c_{1}c_{2}}\cdots s_{c_{i-1}c_{i}}(x) \in \l_{c_i}$ for $i=2,\cdots, n$ which maps a point in $\l_{c_1}$ to a point in $\l_{c_i}$, and these two links are identified under the gluing maps. We can then use repeatedly the previous identity along a path  $\gamma=(c_{1}c_{2}\cdots c_{n})$.
If we define $H^{\l}_{c_{1}c_{i}}\equiv h_{c_{1}c_{2}}\cdots h_{c_{i-1}c_{i}}$, $H^{\l}_{c_{i}c_{1}}$ its inverse  and $\bm{A}^{\l}_{c_{1}c_{i}}\equiv \sum_{k=2}^{i} H^{\l}_{c_{k}c_{1}} a_{c_{k-1}c_{k}}H^{\l}_{c_{1}c_{k}} $
we get:
\be
\bm{z}^{c_{i}}({S_{c_{1}c_{i}}(x)})= H^{\l}_{c_{i}c_{1}}\bm{z}^{c_{1}}({x}) H^{\l}_{c_{1}c_{i}} + \bm{A}^{\l}_{c_{1}c_{i}}.
% \qquad x\in f_{{c}_{1}}\cap \cdots \cap f_{{c}_{n}}\equiv \l.
\ee
If we take the path to form a closed loop ($c_1=c_i$), we get that for any edge $\l_c$ and
every point $x \in \l_c$ there exist gluing maps $S_{c}^{\l}$, holonomies $H^{\l}_{c}$ and translations $ A_{\l}^{c}$ such that:
\be\label{heq}
\bm{z}^{c }({S^{\l}_{c}(x)})= H^{\l}_{c}\bm{z}^{c}({x}) H^{\l}_{c} + \bm{A}^{\l}_{c}.
\ee

Let us first  choose a parametrization of the edge $\l_c$ in terms of a parameter $\varphi$ proportional to
the length from a given point. Since the induced metric on the edge is given by $ \rd s^{2} = (\rd \bm{z^{c}})^{2}= \omega^{2} \rd \varphi^{2}$, $S^{\l}_{c}(\varphi)$ has to be a isometry of the edge, and therefore it is necessarily a translation in the length parametrization: $ S^{\l}_{c}(\varphi) = \varphi + \theta^{\l}_{c}$, where $\theta^{\l}_{c} $ is a fixed angle.

What is remarkable is that the helix is a solution of (\ref{heq}), where $H^{\l}_{c}$ is a rotation of angle $\theta^{\l}$
around the axis $\hat{\w}^{c}_{\l}$, and the translational component is $\bm{A}^{\l}_{c}= K \theta^\l \hat{\w}$. Note that since the angle of rotation is the same for any holonomy $H^\l_c$ which loops once around the link $\l$ (and only that link), we do not need a cell-label on the angle $\theta^\l$.
From  the analysis done in three-dimensional gravity \cite{3d} we know that we can interpret
$H^{\l}_{c}$ as the discrete curvature, while $\bm{A}^{\l}_{c}$ represents a discrete torsion.
We see that if the geometry is non-Regge then the torsion does not vanish. This was already conjectured in \cite{LSpeziale1, FGZ}.

We have arrived again at the conclusion that each link of a spinning geoemtry is the form of a helix. This shows that given a set of twisted geometry data $(\bm{X}_{cc'},h_{cc'})$, we can represent it in terms of a spinning geometry $(\D_\l^{c},\J_{\l}^{c})$ as long as we chose $\hat{\w}_\l^{c}$ to be parallel to the axis of rotation of $H^{\l}_{c}$.

\section{Discussion}
The $\SU(2)$-gauge invariant phase space of loop gravity $P_{\Gamma}$ can be represented by either twisted geometries or spinning geometries, which are both generalizations of Regge geometries. Twisted geometries have the advantage that each cell takes the form of a flat polyhedron, but the disadvantage that they do not form continuous geometries. Spinning geometries are continuous at the price of having cells with curved boundaries.

We studied spinning geometries with an aim to choose a specific form for the links to reduce the ambiguity in the shape of cells. We found that minimizing the link lengths while keeping the nodes fixed results in each link taking the form of a helix. The cells of spinning geometries are similar to polyhedra, except that the links are helical rather than straight.  We also found that helices appear necessarily from the consistency of the gluing
determined by the holonomies. Allowing for the links to be helical accounts for the extra degrees of freedom in the loop gravity phase space which are not present in a Regge geometry.

Our analysis shows that any given spinning geometry corresponds to a point in the phase space $P_{\Gamma}$, parameterized in terms of holonomies and fluxes.
The axis of the product of holonomies which loop around a link $\l$ is parallel to the axis of the helix, and the angle is encoded into the composition of gluing maps along the link.
The question that we haven't resolved yet  is whether we can always reverse this statement and
construct a closed network  of spinning helices for arbitrary holonomy data.
We believe this to be very likely but a proof is necessary.

Another question that needs a deeper understanding is whether we can reconstruct the geometry of the faces from the requirement that the boundary edges are helices and that the normal components of the frame fields are continuous.
One natural conjecture is that the shapes of the faces have to be minimal surfaces. This is what happens for bubbles separating two domains of equal pressure due to the Laplace equation \cite{bubbles}. 

An interesting outcome of this analysis is the appearance of non-trivial torsion along the edges of the cells.
The connection $\A$ in question is the Ashtekar-Barbero connection, related to the spin connection\footnote{The spin connection $\Gamma$ is the metric-compatible connection determined entirely from the metric.} by $\A=\Gamma + \K$ where $\K$ is the extrinsic curvature. The torsion of $\A$ is therefore a measure of $\K$.
However, one usually expects this torsion to be supported on the faces and come entirely from the holonomies.
Here we see that on top of the holonomy contribution we have an additional contribution supported on the edges. This seems to be an intrinsic property of the three-geometry that is not related to the extrinsic curvature. A deeper understanding of this is clearly needed.

Finally one of the most exciting outcomes of this work is that it may give new insights into how to formulate a consistent dynamics on the discrete geometry of loop gravity. If we could assign a continuous connection and triad to a given set of loop gravity data, we would be able to write the continuous scalar constraint\footnote{The scalar constraint in loop gravity is the generator of dynamics.} in terms of these fields. The piecewise-flat and piecewise-torsionless nature of spinning geometries would reduce the scalar constraint to a more simple form which is supported only on the two-skeleton, which may help to write a dynamics in terms of holonomies and fluxes. If successful, this would provide for the first time an anomaly-free means to relate the dynamics of loop gravity with the dynamics of general relativity.

\acknowledgements
The authors would like to thank James Ryan, Eugenio Bianchi and Biancha Dittrich for
discussions on this project. This research was supported in part by Perimeter Institute for Theoretical Physics. Research at Perimeter Institute is supported by the Government of Canada through Industry Canada and by the Province of Ontario through the Ministry of Research and Innovation. This research was also partly supported by grants from NSERC.

\end{document}